# High-Q $Si_3N_4$ microresonators based on a subtractive processing for Kerr nonlinear optics

Zhichao Ye,* Krishna Twayana, Peter A. Andrekson, and Victor Torres-Company

*Department of Microtechnology and Nanoscience(MC2),Photonics Laboratory, Chalmers University of Technology, SE-41296, Sweden*
*zhichao@chalmers.se*

**Abstract:** Microresonator frequency combs (microcombs) are enabling new applications in frequency synthesis and metrology – from high-speed laser ranging to coherent optical communications. One critical parameter that dictates the performance of the microcomb is the optical quality factor (Q) of the microresonator. Microresonators fabricated in planar structures such as silicon nitride ($Si_3N_4$) allow for dispersion engineering and the possibility to monolithically integrate the microcomb with other photonic devices. However, the relatively large refractive index contrast and the tight optical confinement required for dispersion engineering make it challenging to attain $Si_3N_4$ microresonators with Qs > $10^7$ using standard *subtractive* processing methods – i.e. photonic devices are patterned directly on the as-deposited $Si_3N_4$ film. In this work, we achieve ultra-smooth $Si_3N_4$ microresonators featuring mean intrinsic Qs around 11 million. The cross-section geometry can be precisely engineered in the telecommunications band to achieve either normal or anomalous dispersion, and we demonstrate the generation of mode-locked dark-pulse Kerr combs as well as soliton microcombs. Such high-Qs allow us to generate soliton microcombs with photodetectable repetition rates, demonstrated here for the first time in $Si_3N_4$ microresonators fabricated using a subtractive processing method. These results enhance the possibilities for co-integration of microcombs with high-performance photonic devices, such as narrow-linewidth external-cavity diode lasers, ultra-narrow filters and demultiplexers.

## 1. Introduction

Silicon nitride ($Si_3N_4$) is a dielectric material that can be fabricated with standard CMOS processing tools. It has a large transparency window that extends into the visible range and therefore complements the suite of applications enabled by silicon photonics [1]. One of the most prominent applications is in the field of nonlinear optics [2], enabled by the possibility to precisely engineer the dispersion and achieve high optical confinement in a single-core geometry. Recent demonstrations include multi-octave supercontinuum generation [3,4] and carrier-offset-frequency stabilization of mode-locked lasers [5], ultrabroadband frequency conversion [6] and octave-spanning microresonator frequency combs [7,8].

The strong confinement and the relatively large refractive index difference with the cladding make the waveguide very susceptibility to scattering losses arising from nanometer-level roughness at the interfaces. In single-core $Si_3N_4$ microresonators, this results into a practical tradeoff between dispersion engineering and achievable quality factor (Q), which has dramatic consequences particularly in the generation of microresonator frequency combs. For instance, high-Q factors are required to lower the power to operate soliton microcombs in large-mode-volume microresonators. This is of practical relevance for co-integrating pump laser diodes with $Si_3N_4$ microcombs [9] operating at photodectable (< 100 GHz) repetition rates [10].

Record mean Qs in the order of 15 million in high-confinement $Si_3N_4$ with dispersion-engineered structures have been recently achieved with the Damascene reflow process [11]. The waveguide cross-section can be optimized for low dispersion over an ultra-broadband bandwidth [3], and soliton microcombs with photodetectable repetition rates have been reported [11]. In this technique, the $Si_3N_4$ film is deposited on a pre-patterned silica preform [12,13] that has been previously reflowed. This process results into a significantly decreased sidewall roughness [14] but has the detrimental effect of limiting the uniformity of waveguide height across the wafer due to an aspect ratio dependent etching [15,16].

Traditionally, $Si_3N_4$ waveguide cores are defined via a subtractive method, where the $Si_3N_4$ film is lithographically patterned upon deposition on an oxidized silicon wafer [17–20]. Typically, uniformity of film thickness ~ 2% across a wafer can be achieved by low-pressure chemical vapor deposition (LPCVD) of $Si_3N_4$. The issue with the etching susceptibility to the aspect ratio can be easily overcome by allowing extra etching time during dry etching. Extremely high Qs have been reported via subtractive processing [18,21,22], but only for isolated resonances or for geometries that are not optimized in terms of dispersion. The subtractive process is believed to be severely impaired by the sidewall roughness caused by chemical etching, and up to now it has been unclear whether high Qs across the entire spectrum could be obtained in a reproducible manner. Here, we defeat the tradeoff between optical confinement and limited Qs, and demonstrate dispersion-engineered $Si_3N_4$ microresonators with mean Qs ~ 11 million, i.e. comparable to those attained by the Damascene reflow process [11] but using a subtractive method instead. We show the possibility to achieve both normal and anomalous dispersion in high-confinement $Si_3N_4$ microresonators, leading to mode-locked dark-pulse Kerr combs [23] or soliton microcombs [24]. We also demonstrate a soliton microcomb operating at 100 GHz. This technology is also compatible with high-speed thermo-optic tuning [25], based on which we show broadband soliton microcombs without tuning the pump laser.

The paper is organized as follows. In Section II, we describe our fabrication process for $Si_3N_4$ microring resonators, and we present characterization of the top surface roughness of the $Si_3N_4$ film and sidewall roughness of fabricated waveguides. In section III, we focus on measurements of dispersion and Qs for our microring resonators. In section IV, we show both mode-locked dark pulse Kerr combs and soliton microcomb generation from our dispersion-engineered microring resonators.

## 2. Fabrication flow and characterization of $Si_3N_4$ thin film and waveguides

We start our fabrication process from a 3-inch Si wafer with 3 µm thermally oxidized $SiO_2$. The fabrication flow is shown in Fig 1.(a). $Si_3N_4$ films with thicknesses above 600 nm are selected for waveguides with normal and anomalous dispersion. Crack barriers [20] and thermal cycling process [17] are adapted to overcome the crack formation due to high tensile stress in these relatively thick $Si_3N_4$ films. S1813 resist and direct laser writer are used to define the patterns for the crack barriers, and a buffered oxide etch is used to etch into $SiO_2$, which eventually forms 3 µm deep trenches that are used to terminate crack formation originated from the edge of the wafer. The first $Si_3N_4$ film layer with thickness of 350 nm is deposited in an LPCVD furnace. Then, the $Si_3N_4$ thin film is annealed at 1100 ºC under $N_2$ ambient for 3 hours [26]. Standard cleaning is applied to avoid potential contamination from the high-temperature furnace. We noticed that a clear interface between two layers can be seen under SEM after dry etching $Si_3N_4$ waveguide if we skip the standard cleaning process prior to the deposition of second layer. In the end, the second layer of $Si_3N_4$ thin film is deposited to achieve a target thickness of either ~ 600 nm or 740 nm in this work. Fig 1.(b) shows an SEM image of a fully etched $Si_3N_4$ waveguide indicating a continuous interface between the two layers. Atomic-force microscopy (AFM) is used to measure the top surface roughness of the as-deposited $Si_3N_4$ film with thickness ~ 600 nm. The AFM result is shown in Fig 1.(c), with a root mean square (RMS) roughness of our as-deposited thin film ~ 0.18 nm, which is half of the RMS surface roughness reported in [21] before chemical mechanical polishing.

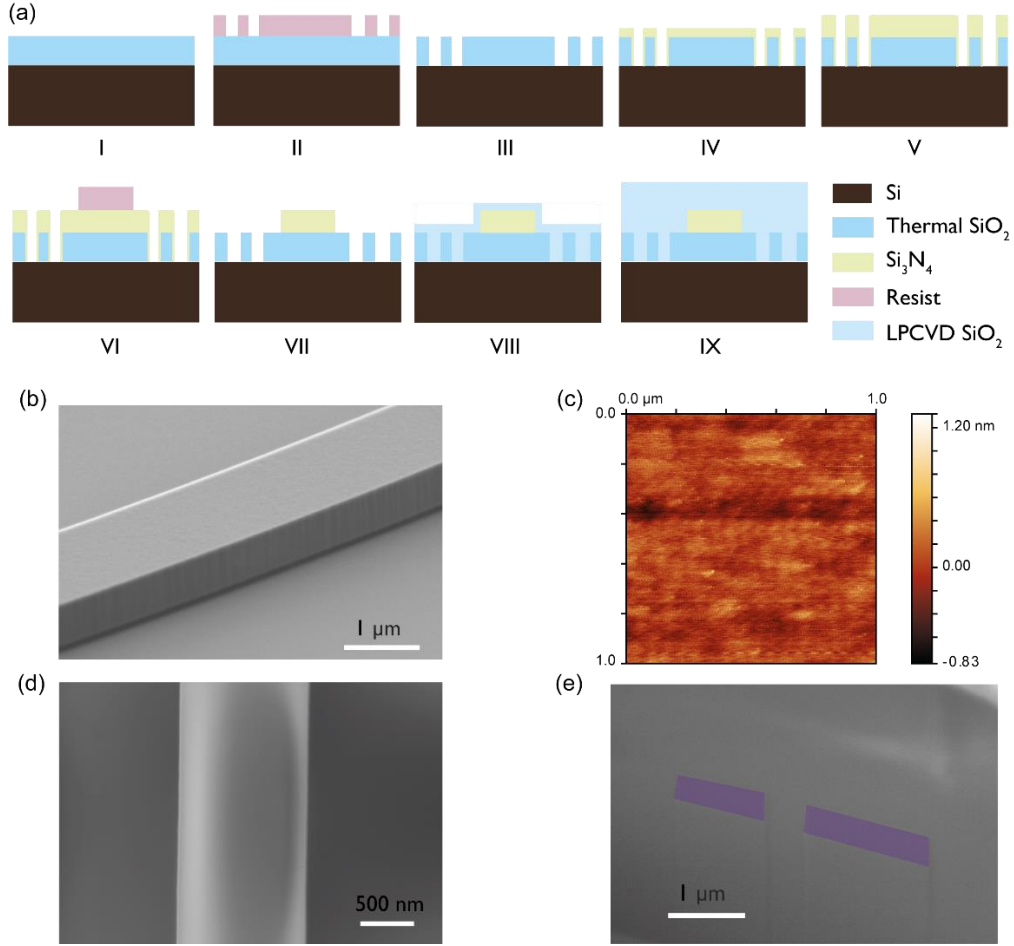

Fig. 1. (a) Fabrication flow of $Si_3N_4$ waveguides starting from deposition of crack-free thin film until overclad of $Si_3N_4$ waveguide. (b) Perspective view SEM image of fabricated $Si_3N_4$ waveguide. Little roughness can be observed, and no clear interface between two $Si_3N_4$ layers can be seen. (c) AFM measurement of the top surface of $Si_3N_4$ with thickness ~ 600 nm. (d) Top view SEM image of fabricated $Si_3N_4$ waveguide. (e) SEM image of coupling region milled by FIB. The $Si_3N_4$ bus and ring waveguides are painted with purple color. The gap between ring and bus waveguides is ~ 400nm, and no air void has been observed.

After $Si_3N_4$ thin film deposition, electron beam lithography using MaN 2405 resist is adapted to pattern $Si_3N_4$ microring resonators. MaN 2405 resist with thickness ~700nm is exposed by Raith 5200 using a beam step size of 2 nm and beam current of 1.8 nA in order to deliver a dose ~ 600 µC/cm². $Si_3N_4$ is dry etched by inductively coupled plasma reactive ion etching using $CHF_3$ and $O_2$ based etchants. The dry etching recipe is carefully optimized to achieve $Si_3N_4$ waveguides with minimum sidewall roughness. The sidewall angle of the waveguide is ~ 85 degrees. Figure 1. (d) shows a top view SEM image. We use ProSEM software to evaluate the line edge roughness of our $Si_3N_4$ waveguide, and we get ~ 1 nm RMS roughness (approaching the limitation of SEM) and 450 nm for correlation length. The perspective view SEM image of the dry etched waveguide is shown in Fig 1.(b), where a small sidewall roughness can be appreciated. After dry etching process, the $Si_3N_4$ waveguide is annealed at 1100 ºC under $N_2$ ambient for 3 hours to outgas hydrogen. TEOS $SiO_2$ with thickness ~ 500 nm is deposited under low pressure (~ 200 mTorr) and high temperature (~ 800 ºC), which contribute to conformal deposition. Then TEOS $SiO_2$ is annealed under 1100 ºC under $N_2$ ambient to densify TEOS $SiO_2$. We noticed ~ 5% shrinkage of film thickness and an increase of refractive index for TEOS $SiO_2$ after annealing process. Focused ion beam (FIB) milling is applied at the coupling region of the microring resonator. The SEM image of the coupling region after FIB milling is shown in Fig.1(e), where no air void can be observed between the bus and ring waveguide. Finally, ~ 2 µm plasma enhanced chemical vapor deposition (PECVD) $SiO_2$ is deposited to clad the devices.

### 3. Characterization of microring resonators

In this section, we characterize the Qs and dispersion of our fabricated microring resonators. Since precise frequency calibration is required to measure the Q value and dispersion of microring resonators, we use a self-referenced fiber frequency comb (MenloSystems FC-1500 with repetition rate of 250 MHz) to calibrate the frequency of our laser when we sweep it from 1520 nm to 1620 nm [27].

We first characterize our microring resonators having a ring waveguide geometry of height 600 nm and width 1850 nm. The Qs and exact wavelength of all the resonances within the measured spectrum are extracted. The integrated dispersion (defined as $D_{int} = \omega_\mu - \omega_0 - \mu D_1 = D_2 \mu^2/2 + D_3 \mu^3/6 + \ldots$ where $\omega_\mu$ is the angular frequency of the µ-th resonance relative to the reference resonance $\omega_0$, and $D_1/2\pi$ is the FSR [24]) from the fundamental TE mode family is shown in Fig 2.(a). The retrieved values from the fitting are $D_1/2\pi$ ~ 105.2 GHz, $D_2/2\pi$ ~ -0.75 MHz and $D_3/2\pi$ ~ 2.8 kHz. The converted $\beta_2$ is 76 ± 2 $ps^2/km$ which is close to our simulation result of 90 $ps^2/km$ at 1570 nm. Two avoided mode crossings are observed at 1540 nm and 1560 nm which shall be used to initialize dark-pulse Kerr combs in section IV. Figure 2 (b) shows the intrinsic Q of all the resonances within the measurement spectrum. The mean intrinsic Q is 11.4 ×$10^6$ which corresponds to an equivalent loss ~3 dB/m. In order to have more statistics, the histogram of intrinsic linewidth of 6 microring resonators with different gaps coming from the same chip is shown in Fig. 2(c). The highest probable intrinsic linewidth is between 15-18 MHz. A representative critically coupled resonance is shown in Fig. 2(d). The fitting parameters result in a full width half maximum (FWHM) of 23.4 MHz and intrinsic linewidth of 11.7 MHz ($Q_i$ ~ 16×$10^6$).

We also fabricated microring resonators with anomalous dispersion. The width and height of the ring waveguide are 2000 nm and 740 nm, respectively. The bus waveguide cross-section has an identical design to the ring waveguide in order to achieve high coupling ideality [28]. The integrated dispersion of the fundamental TE mode family is shown in Fig. 2(e). The FSR is 100 GHz and the converted $\beta_2$ is -67 ± 1 $ps^2/km$ at 1570 nm which is close to our simulation result of -50 $ps^2/km$. Moreover, the attained $D_3/2\pi$ ~ 0.4 kHz indicates a negligible $\beta_3$ which is necessary for bright soliton comb generation with broad spectrum [29]. Four avoided mode crossings are observed at 1545nm, 1566nm, 1586nm and 1607 nm, but they represent less than 100 MHz deviation from the fitted third-order polynomial curve. The intrinsic Q for all the resonances is shown in Fig.2 (f), resulting in a mean value of 12.5 × $10^6$. The histogram of intrinsic linewidth is shown in Fig.2 (g), where the highest probable intrinsic linewidth is around 15 MHz. A representative resonance with near critical coupling is shown in Fig.2 (h). The fitting leads to a FWHM of 19.8 MHz and intrinsic linewidth ~ 11.7 MHz.

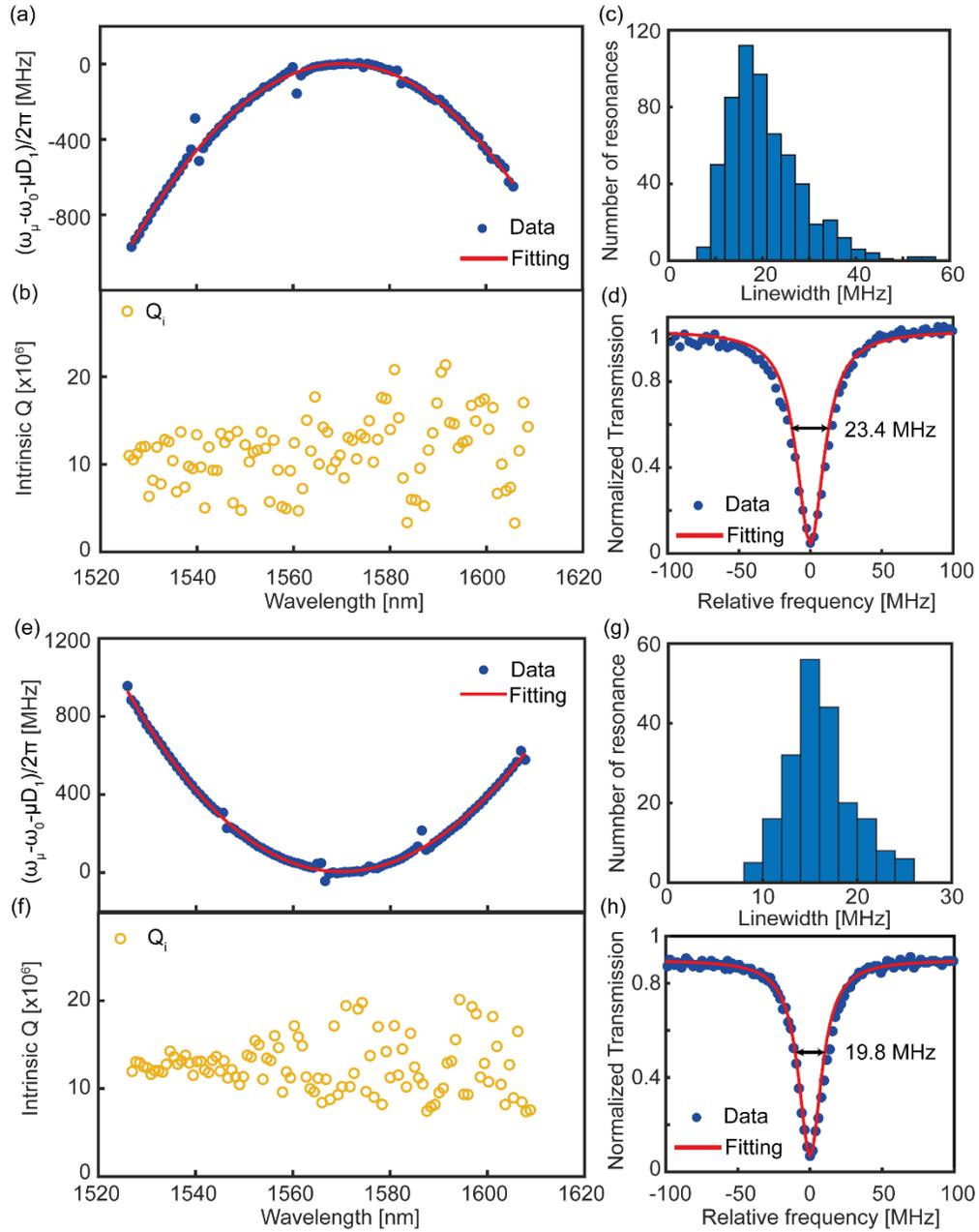

Fig. 2. Intrinsic Q and dispersion characterization of microring resonator with normal (height 600 nm × width 1850 nm, a, b, c, d) and anomalous dispersion (height 740 nm × width 2000 nm, e, f, g, h). (a) Integrated dispersion of fundamental TE mode. Two weak mode crossings are observed. (b) Intrinsic Q of resonance from fundamental TE mode family. (c) Histogram of intrinsic linewidth from 6 microring resonators. (d) A representative resonance with critical coupling and FWHM of 23.4 MHz. (e)-(g) idem to (a)-(d) but for the anomalous dispersion design.

## 4. Soliton frequency comb generation

In this section, we present soliton comb generation from our high Q microring resonators. Firstly, we investigate one of our microring resonators with normal dispersion. Since mode crossing is beneficial to initialize dark-pulse Kerr combs [23], we pump our device at 1540 nm (see integrated dispersion in Fig.2 (a)) with ~200 mW on-chip power. We slowly tune the wavelength of the pump laser from the blue side to approach the resonance and eventually obtain the comb shown in Fig.3 (a) whose envelope, high conversion efficiency (23 % in this case [30], calculated by the sum of comb power, excluding the pump, normalized to the pump power) and low noise radio-frequency spectrum (see Fig. 3(b)) lead us to conclude that this is a mode-locked dark-pulse Kerr comb with 2 FSR repetition rate.

We now turn our attention to a microring resonator with large FSR ~ 0.9 THz featuring anomalous dispersion. The microring has a radius of 25 µm, and the height and width of the ring waveguide are 740 nm and 1500 nm. The bus waveguide with height 740 nm and width 800 nm result in a single mode which is designed for high coupling ideality [28]. The gap between bus and ring waveguide is 700 nm which results in overcoupling, and the loaded Q of this device is ~ $1 \times 10^6$ at 1540 nm. On chip power ~ 125 mW is used to pump the device at 1540 nm. We obtained a multi-soliton comb by directly tuning the pump laser wavelength from the blue to the red side of the resonance, and monitored the comb power as shown in Fig.3 (d). Then, the backward tuning method [31] was used to achieve a single soliton comb which is shown in Fig.3 (c). The obtained soliton comb covers a spectrum from 1300 nm to 2200 nm with a clear dispersive wave generated at 2100 nm [32].

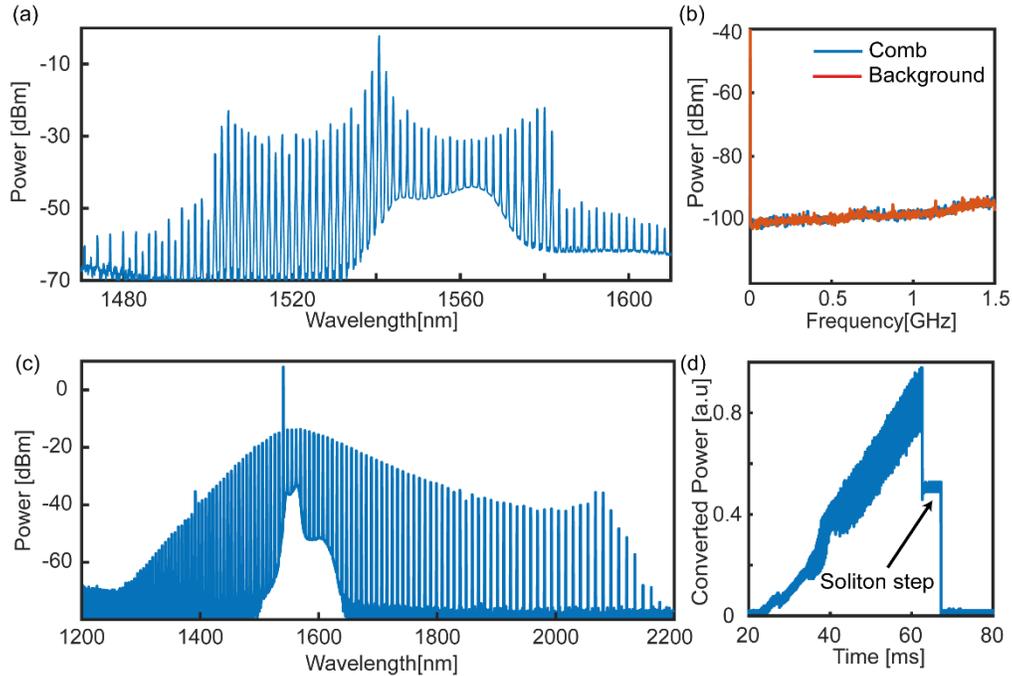

Fig. 3. Microcomb generation from microring resonators with line spacing 210 GHz and 0.9 THz. (a) Mode-locked dark-pulse Kerr comb generation from 105 GHz microring. Its RF spectrum is shown in (b). (c) Soliton microcomb generation from 0.9 THz microring. (d) Soliton step in converted power trace by sweeping the pump laser across the resonance with on-chip power ~ 125 mW.

This platform renders suitable for the co-integration of microresonators with thermo-optic heaters [25]. This allows to pump the microresonator device with a fixed continuous-wave laser and tune the resonance instead. We investigated a 100 GHz microring resonator with anomalous dispersion. We pump our device at 1559 nm with an on-chip power ~100 mW. The heater signal to generate a soliton microcomb is shown in Fig. 4(a), and the corresponding converted power trace is shown in Fig. 4(b) with a zoomed in displayed in Fig. 4(c). The small voltage kick at the end of microheater tuning around 1.7 ms helps us to stop the thermal tuning quickly and provide small backward tuning to stabilize the soliton comb. The obtained single soliton comb initialized by the microheater is shown in Fig.4 (d). A broad spectrum ranging from 1470 nm to 1700 nm with a smooth envelope is observed, without severe distortions due to avoided mode crossings. The conversion efficiency is only 0.8% due to the naturally small duty cycle of the circulating soliton pulse [33]. The RF spectrum in Fig. 4(e) gives indication of mode-locking. We believe this is the first time a soliton microcomb with a photodetectable repetition rate is reported using a $Si_3N_4$ microresonator fabricated using a subtractive process.

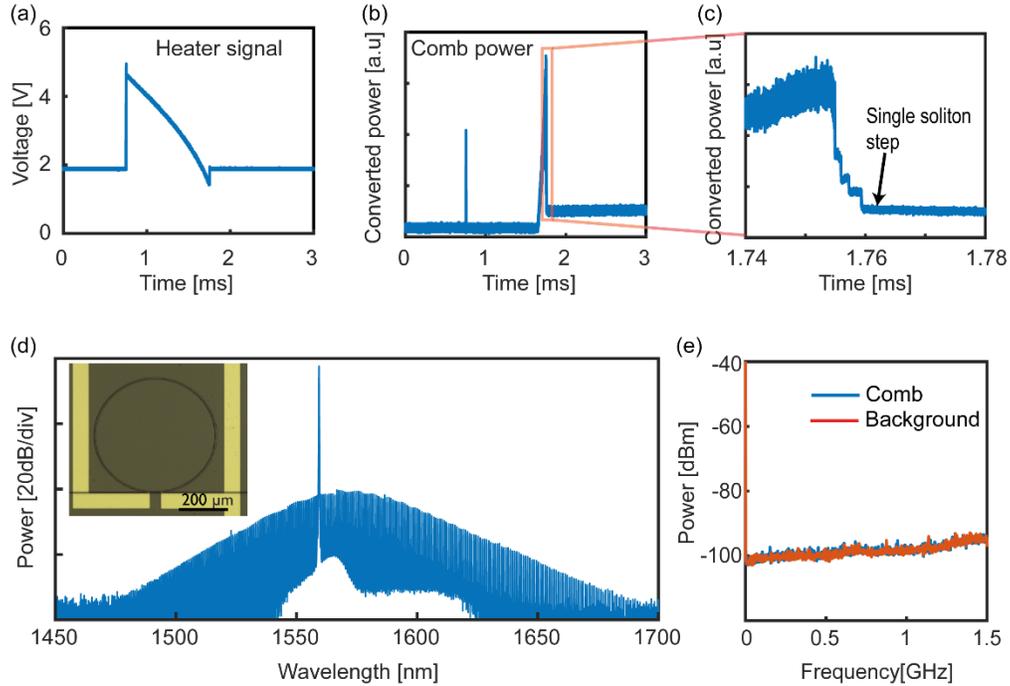

Fig. 4. Soliton comb generation from microring resonator with FSR 100 GHz with fixed pump laser and on-chip thermal heater. (a) The signal for microheater to generate a soliton microcomb. The corresponding converted power trace in shown in (b). (c) Zoomed in trace in (b) showing single soliton step. (d) Single soliton comb generated with on-chip power ~ 100 mW. The microscope image of microheater is inset. The corresponding RF spectrum is shown in (e).

## 5. Conclusion

We have presented high Q (mean intrinsic Qs up to 11 million) dispersion-engineered $Si_3N_4$ microring resonators fabricated using an optimized subtractive processing method. This fabrication technique overcomes issues associated with an etching dependent aspect ratio, allowing for better device homogeneity across the wafer. We have demonstrated both dark-pulse Kerr combs and soliton microcombs, and demonstrated a soliton microcomb operating at 100 GHz repetition rate. The raw data of the measurement results within this work will be released in Zenodo upon publication.

### Funding

This project is funded by the Swedish Research Council (VR, 2015-00535, 2016-03960 and 2016-06077) and the European Research Council (ERC Consolidator Grant, GA 771410).

### Acknowledgements

The authors thank Michael Alexander Bergmann for AFM measurement and the cleanroom staff from Chalmers nanofabrication laboratory for fruitful discussions.

**Disclosures**

The authors declare no conflict of interest.